\documentclass[12pt]{article}

\newcommand{\eq}{\begin{equation}}
\newcommand{\eqx}{\end{equation}}
\newcommand{\eqn}{\begin{eqnarray}}
\newcommand{\eqnx}{\end{eqnarray}}
\newcommand{\f}[2]{\frac{#1}{#2}}
\newcommand{\lm}{\lambda}

\newcommand{\tr}{\mbox{\rm tr}}

\newcommand{\om}{\omega}
\newcommand{\sg}{\sigma}

\newcommand{\dl}{\delta}
\renewcommand{\AA}{{\cal A}}
\newcommand{\TT}{{\cal T}}
\newcommand{\AAd}{{\cal A}^\dagger}

\newcommand{\OM}{\Omega}
\renewcommand{\SS}{{\cal S}}

\newcommand{\arr}[4]{%
\left(\begin{tabular}{c|c}
$#1$ & $#2$ \\
\hline
$#3$ & $#4$
\end{tabular}\right)}

\renewcommand{\det}{\mbox{\rm det}}

\title{Wishart and Anti-Wishart random matrices}

\author{Romuald A. Janik$^{a,b}$, Maciej A. Nowak$^b$\footnote{
e-mail: {\tt janik@nbi.dk}, {\tt nowak@alphas.if.uj.edu.pl}}\\ \\
$^a$ The Niels Bohr Institute,\\
Blegdamsvej 17, DK-2100 Copenhagen,\\ 
Denmark\\
$^b$ M. Smoluchowski Institute of Physics\\ 
Jagellonian University\\ 
Reymonta 4, 30-059 Cracow, Poland}

\begin{document}

\maketitle

\begin{abstract}
We provide a compact exact 
representation for the distribution of the 
matrix elements of the Wishart-type random matrices $\AAd \AA$, 
for any finite number of rows and columns of $\AA$, 
without any large $N$ approximations.
In particular we treat the case when the Wishart-type random matrix
contains redundant, non-random information, which is a new result.
This representation is of interest for a procedure
of reconstructing the redundant information hidden in Wishart matrices, 
with potential applications to numerous models based on  biological,
social and artificial intelligence  networks.   
\end{abstract}

\section{Introduction}

Random matrices of the form $\OM=\AAd\AA$, where $\AA$ is a random
rectangular matrix of size $n\times k$ occur in many
applications. Introduced in the classical paper by
Wishart~\cite{WISHART}, (probably the first application of random
matrix models), they form the cornerstone of multivariate
statistical analysis~\cite{WILKS}. 
Standard applications include biology, economy, telecommunication, to
mention a few. Usually, $n$ samples of $k$-dimensional rows of data are used
to  construct the covariance matrix. Typical measurements
are e.g. samples of large number of meteorological observations
at various sites collected at time intervals~\cite{PREISEN}, 
high frequency financial 
data for large portfolios~\cite{BOUCHAUD} or
  wireless channels with multiple antennae or receivers~\cite{COCKTAIL}. 
 Wishart matrices appear also in fundamental science,
ranging from 
condensed matter physics~\cite{PICHARD}, nuclear physics~\cite{WEIDENMULLER}
to  chiral quantum chromodynamics~\cite{VERSHURZAH}.
Recently, the topic of retrieving the redundant information from
Wishart matrices got a new
twist due to the spectacular increase in computing and storage powers.
The key problem is to handle effectively 
large incidence matrices constructed by Intelligent Retrieval (IR) 
engines.  
These  new  challenges  vary from eliminating the redundant information
in Internet traffic, through exploiting the hidden information 
of knowledge  networks~\cite{ZHANG} to  unraveling the cross-correlations in 
bioinformatics and genetics.

Let us consider a matrix $\Omega=\AAd\AA$. 
There are two qualitatively different situations that may appear. When
$n\geq k$, all the elements of $\OM$ are indeed random. Their
probability distribution was first derived by Wishart in~\cite{WISHART}.
In the opposite case $n<k$, there is a lot of redundancy in $\OM$,
and only a part of the matrix elements are random, the remaining elements
are unambiguously determined in terms of the random ones. 
The analytical expression for the probability distribution of
matrix elements of $\OM$ in this `Anti-Wishart' case (this name was 
coined by Yi-Cheng Zhang) remained,
however, unknown. The numerical study of the Anti-Wishart case was  
presented in~\cite{ZHANG}.
In this paper we would like to derive this probability distribution as
well as to provide a procedure of reconstructing the redundant
information from the first $n$ rows of the matrix. 
 All these results are exact for any $n$ and $k$,
without any large $N$ approximations.

The plan of this paper is as follows. In the next two sections we will derive
recurrence relations for the probability distributions for {\em complex}
Gaussian matrices $\AA$. In section 4 we will use these relations to
rederive the classical Wishart case, and then proceed to analyze the
new Anti-Wishart case. For completeness, in section 5 we will state
some known results for the joint
eigenvalue distributions in the Anti-Wishart case.
 We will end the paper with a discussion and
two appendices  which state the analogous results, when $\AA$ are {\em
real} Gaussian matrices and include some mathematical details.

\section{Wishart and Anti-Wishart random matrices}

Let $\AA$ be a {\em complex} rectangular matrix of size $n\times k$
taken from a Gaussian ensemble:
\eq
P(\AA)=\f{1}{\pi^{nk}} e^{-\tr \AAd\AA}
\eqx
We would like to derive the probability distribution of the elements
of the Wishart matrix $\OM=\AAd\AA$. We thus have to evaluate
\eq
P_{n,k}(\OM)=\int d\AA d\AAd \,\dl(\OM-\AAd\AA)\, P(\AA)
\eqx
When $n\geq k$ (`Wishart case') the resulting distribution was
obtained by Wishart after a quite intricate calculation for real
$\AA$ (see e.g. \cite{WILKS}). The analogous result for complex matrices is
\eq
P_{n,k}^{WISHART}(\OM)=C_{n,k}\left( \det\, \OM \right)^{n-k} e^{-\tr\, \OM}
\eqx
where $C_{n,k}$ is a normalization constant.

When $n<k$ (`Anti-Wishart case') no such explicit formula was known. 
The goal of this work is to give a simple unified derivation of
$P_{n,k}(\OM)$ which works in both cases and to provide a procedure of
reconstructing the redundant information from the first $n$ rows.
The problem of reconstructing the $\OM$ matrix from more realistic
sparse data will be considered in a subsequent work.

As was noted in \cite{ZHANG},
the determination of $P_{n,k}(\OM)$ can be easily translated, using the
integral representation
\eq
\dl(\OM-\AAd\AA)=\f{1}{(2\pi)^{k^2}}\int d\TT e^{i \tr\, \TT (\OM-\AAd\AA)}
\eqx
into the integral
\eq
\label{e.tint}
P_{n,k}(\OM)=C'_{n,k}\int d\TT \det(1+i\TT)^{-n} \, e^{i\, \tr\, \TT\OM}
\eqx
where $\TT$ is a $k\times k$ {\em hermitian} matrix.

\section{Recurrence relations}

The integrals of the form (\ref{e.tint}) with $\OM$ diagonal were
considered for $n\geq k$ in a beautiful paper of Fyodorov
\cite{FYODOROV}. 
Here we will slightly generalize his procedure for arbitrary
non-diagonal $\OM$. Although the integrals (\ref{e.tint}) are invariant
with respect to unitary transformations, this generalization is
necessary in the case $n<k$. Indeed, then the matrix $\OM$ has $k-n$
{\em exact} zero 
eigenvalues and the Jacobian for the diagonalization
$\OM=U\Lambda U^\dagger$ will be quite nontrivial.

Let us first decompose the matrices 
$\TT \equiv \TT_k$ and $\OM \equiv \OM_k$ as
\eq
\label{e.decomp}
\TT_k = \arr{t_{11}}{t^\dagger}{t}{\TT_{k-1}} \quad\quad\quad\quad
\OM_k = \arr{\om_{11}}{\om^\dagger}{\om}{\OM_{k-1}}
\eqx
We will derive a recurrence relation by first integrating over $t_{11}$ and
then over the vector $t$. To this end we use the identity
\eq
\det(1+i\TT_k) = \left(1+it_{11} +t^\dagger (1+i\TT_{k-1})^{-1} t \right)
\cdot \det(1+i\TT_{k-1})
\eqx 
The integral over $t_{11}$ can be done by residues giving
\eq
\label{e.partial}
\int d\TT_{k-1} e^{i\tr \TT_{k-1}\OM_{k-1}} (\det(1+i\TT_{k-1}))^{-n} 
\om_{11}^{n-1}e^{-\om_{11}} \nonumber \\
\cdot \,\,\int dtdt^\dagger e^{-\om_{11} t^\dagger (1+i\TT_{k-1})^{-1} t}
e^{i (t^\dagger \om+\om^\dagger t)}
\eqx
The last integral is Gaussian giving 
\eq
\f{1}{\om_{11}^{k-1}}  \cdot \det(1+i\TT_{k-1}) \cdot
e^{-\f{1}{\om_{11}} \om^\dagger (1+i\TT_{k-1}) \om}
\eqx 
Substituting this back into (\ref{e.partial}) leads to
\eq
\int d\TT_{k-1} (\det(1+i\TT_{k-1}))^{-(n-1)} e^{i\tr \TT_{k-1} 
\left( \OM_{k-1}-\f{1}{\om_{11}}\om \om^\dagger \right)}
\om_{11}^{n-k} e^{-\om_{11}-\f{\om^\dagger \om}{\om_{11}}}
\eqx
Hence we are led to the recurrence relation
\eq
\label{e.rec}
P_{n,k}(\OM_k)= C''_{n,k}\, \om_{11}^{n-k} e^{-\om_{11}-\f{\om^\dagger
\om}{\om_{11}}} P_{n-1,k-1}\left(\OM_{k-1}- \f{1}{\om_{11}}\om
\om^\dagger \right)
\eqx

\section{Probability distributions for elements of $\OM$}

We will now use the recurrence relation (\ref{e.rec}) to determine
$P_{n,k}(\OM)$. Let us first consider the easy `Wishart case' ($n \geq
k$). Then repeated use of (\ref{e.rec}) reduces the problem to 
calculating $P_{n,1}(\OM)$ which is just
\eq
P_{n,1}(\om)=\om^{n-1} e^{-\om}
\eqx
Since in the Wishart case the eigenvalues are generically distinct we
may diagonalize the matrix $\OM$, and then the recurrence relation can
be solved immediately \cite{FYODOROV} to get
\eq
\label{e.wishart}
P_{n,k}^{WISHART}(\OM)=C'''_{n,k}\, (\det\, \OM)^{n-k} e^{-\tr \,\OM}
\eqx

Let us now turn to the more interesting Anti-Wishart case ($n<k$).
Then repeated use of the recurrence relation reduces to the initial
condition
\eq
\label{e.delta}
P_{0,k}(\OM_k)=\dl(\OM_k)
\eqx
$\dl(\OM_k)$ is defined on the space of hermitian matrices through the
integral
\eq
\label{e.deltadef}
\dl(\OM_k) = \int d{\cal A} e^{i \tr \,\, \OM_k {\cal A}}
\eqx
where the integration domain covers the space of {\em hermitian}
$k\times k$ matrices.

In the first nontrivial case we then have
\eq
\label{e.awone}
P_{1,k}(\OM_k)=\om_{11}^{1-k} e^{-\om_{11}-\f{\om^\dagger
\om}{\om_{11}}} \dl\left(\OM_{k-1}- \f{1}{\om_{11}}\om
\om^\dagger \right)
\eqx
Due to the form of the Dirac delta function we may immediately
recognize that the argument of the exponent is just the ordinary trace
of $\OM_k$. 

\subsection*{Solution of the recurrence relations}

We will now obtain an explicit expression for general $n<k$.
The recurrence relation (\ref{e.rec}) expresses the probability
distribution of a $k\times k$ matrix $\OM_k$ by a probability
distribution of a $k\!-\!1 \times k\!-\!1$ one. The first step 
towards finding the general solution is to obtain an explicit
expression for the elements of the relevant $k\!-\!i \times k\!-\!i$
matrix at the $i^{th}$ step of the recursion. 

We therefore have to solve: 
\eqn
\label{e.matrec}
\OM^{(k)}   &=&\OM \\
\label{e.matreceq}
\OM^{(i-1)} &=&\OM^{(i)}_{i-1} -\f{\om^{(i)}\om^{(i)\dagger}}{\om^{(i)}_{11}}
\eqnx
where the superscript $(i)$ denotes the $k-i$ step of the recursion
while the subscripts are defined through the decomposition analogous
to (\ref{e.decomp}):
\eq
\OM^{(i)} = \arr{\om^{(i)}_{11}}{\om^{(i)\dagger}}{\om^{(i)}}{\OM^{(i)}_{i-1}}
\eqx
The `reduced' matrix $\OM^{(i)}$ is of size $i\times i$.

Once the explicit expressions for $\OM^{(i)}$ are known, the
probability distribution for the Anti-Wishart case can be written as:
\eq
\label{e.partialaw}
P_{n,k}(\OM_k)\propto \left[ \prod_{i=0}^{n-1} \om_{11}^{(k-i)} \right]^{n-k} 
\cdot e^{-\sum_{i=0}^{n-1}
\left\{ \om^{(k-i)}_{11}+\f{\om^{(k-i)\dagger}
\om^{(k-i)}}{\om^{(k-i)}_{11}}
\right\}
} 
\cdot \dl(\OM^{(k-n)})
\eqx 
The matrix-valued delta in (\ref{e.partialaw}) 
of the hermitian $(k-n)\times(k-n)$ matrix $\Omega^{(k-n)}$ is equivalent 
to the set of $(k-n)^2$ deltas for the independent matrix
elements (keeping in mind that the diagonal elements are real and the
off-diagonal ones appear in complex conjugate pairs). 

We will first show that the argument of the exponent is just the trace
of $\OM$. Let us take the trace of both sides of the recurrence
relation (\ref{e.matreceq}):
\eq
\tr\, \OM^{(i-1)} =\tr \, \OM^{(i)} - \om^{(i)}_{11} -
\f{\om^{(i)\dagger}\om^{(i)}}{\om^{(i)}_{11}} 
\eqx
Then the sum in the exponent may be rewritten as
\eq
-\sum_{i=0}^{n-1} \left( \tr \, \OM^{(k-i)} - \tr\, \OM^{(k-i-1)} \right)
\eqx
In this alternating sum all terms cancel except the first and the
last which give $-\tr \, \OM^{(k)}+\tr \, \OM^{(k-n)}$. The last matrix
vanishes due to the Delta function in (\ref{e.partialaw}), while the
first one is just the original matrix $\OM$ (see (\ref{e.matrec})).

In order to deal with the remaining terms in (\ref{e.partialaw}) we
have to explicitly solve the recurrence relations
(\ref{e.matrec})-(\ref{e.matreceq}).

Remarkably enough one can give a compact formula for the reduced
matrices in terms of a ratio of determinants.  
Let us denote by $\OM_{[i]}$ the upper left hand $i\times i$ sub-matrix
of the {\em original} matrix $\OM$:
\eq
\label{e.defi}
\OM=\arr{\OM_{[i]}}{*}{*}{*}
\eqx
Furthermore for each $l,m>i$ we will consider the ($(i+1)\times (i+1)$)
matrix $\OM_{[i],lm}$ obtained by adjoining the $l^{th}$ row and
$m^{th}$ column of $\OM$ to $\OM_{[i]}$:
\eq
\label{e.defii}
\OM_{[i],lm}=\left(
\begin{tabular}{ccc|c}
&          & & $\om_{1m}$ \\
&$\OM_{[i]}$ & & \ldots \\
&          & & $\om_{im}$ \\
\hline
$\om_{l1}$&\ldots &$\om_{li}$ & $\om_{lm}$
\end{tabular}\right)
\eqx
In terms of these data, the solution to (\ref{e.matrec}) can be
expressed through the simple formula
\eq
\label{e.recsol}
\OM^{(k-i)}_{lm}=\f{\det\, \OM_{[i],l+i\,m+i}}{\det\, \OM_{[i]}}
\eqx
We give some details of the proof in Appendix B.
{}From here we can easily read off the elements $\om^{(i)}_{11}$
entering formula (\ref{e.partialaw}):
\eq
\om_{11}^{(k-i)}=\f{\det\, \OM_{[i+1]}}{\det\, \OM_{[i]}}
\eqx
where $\det\, \OM_{[0]}$ is understood as 1.
Putting together the above results, we obtain our final result for
the probability distribution for the Anti-Wishart case:
\eq
P_{n,k}^{ANTI-WISHART}(\OM)=C''''_{n,k}\,
(\det\, \OM_{[n]})^{n-k}\,
e^{-\tr\, \OM}\, 
\prod_{l=n+1}^k \prod_{m=n+1}^k \dl\left( \f{\det\,
\OM_{[n],lm}}{\det\, \OM_{[n]}} \right)  
\eqx
As explained above, the number of independent deltas is $(k-n)^2$. 

The above formula has two important features. The Dirac delta function
shows that part of the matrix elements of $\OM$ are nonrandom and
are expressed deterministically in terms of the first $n$ rows of
$\OM$. In fact this does not depend in any way on the type of
randomness assumed. The reconstruction formula works even for a fixed
$\OM=\AAd\AA$. Moreover the quotient of the determinants is a {\em
linear} function of $\om_{lm}$, thus giving a simple expression
for $\om_{lm}$ in terms of the elements coming from the first $n$ rows
of $\OM$. In addition we note that given a fixed matrix $\OM=\AAd\AA$
empirically, with absolutely no information on $\AA$, we may determine
the size of the $\AA$ matrix by computing the successive reduced
matrices $\OM^{(i)}$, and checking the value of $i$ when these
matrices vanish. 

The pre-factor of the Dirac delta function, which gives
the probability distribution for the first $n$ rows of $\OM$ is linked
with the type of randomness of $\AA$.

\section{Joint eigenvalue distributions}

In this section, for completeness, we give the results for the joint
eigenvalue distributions for Wishart and Anti-Wishart random
matrices, which, in contrast to the probability distributions of the
matrix elements themselves $P_{n,k}(\OM_k)$ derived in the preceding
section, are rather well-known~\cite{EARLY}.
In the Wishart case ($n\geq k$) all the eigenvalues are
generically distinct and the result follows from (\ref{e.wishart}) in
the standard way by including a Vandermonde determinant~\cite{FORRESTER}:
\eq
P^{WISHART}_{n,k}(\lm_1,\ldots,\lm_k)=\prod_i \lm_i^{n-k} e^{-\lm_i}
\cdot \prod_{i<j} (\lm_i-\lm_j)^2 
\label{wisheigfm}
\eqx
In the Anti-Wishart case there are always $k-n$ exact zero-modes. The
remaining $n$ nonzero eigenvalues of $\AAd\AA$ are distributed with
the {\em same} probability distribution as the eigenvalues of
$\AA\AAd$. This can be seen by diagonalizing the rectangular matrix
$\AA$ through $\AA=U \SS V$ with $U\in U(k)$, $V\in U(n)$ and 
\eq
\SS=\left(
\begin{tabular}{c}
$\Lambda$ \\
0
\end{tabular}\right)
\eqx 
is a $k\times n$ matrix with $\Lambda$ diagonal. Then $\AA\AAd= U
\SS\SS^\dagger U^\dagger$ and $\AAd\AA= V^\dagger \SS^\dagger \SS
V$. Hence the nonzero eigenvalues of $\AA\AAd$ and $\AAd\AA$ coincide.

We may then write immediately the joint eigenvalue distribution for
the Anti-Wishart case ($n<k$):
\eq
P^{ANTI-WISHART}_{n,k}(\lm_1,\ldots,\lm_n)=\prod_i \lm_i^{k-n} e^{-\lm_i}
\cdot \prod_{i<j} (\lm_i-\lm_j)^2 
\label{awisheigfm}
\eqx 
We do not discuss here the asymptotic spectral properties of 
(\ref{wisheigfm},\ref{awisheigfm}). For recent results and original references
we refer to recent mathematical literature on this subject~\cite{HAAG}.

\section{Conclusions}

Using the recently advocated properties of the Ingham-Siegel 
\cite{FYODOROV} integrals, 
we have provided  a compact expression for the distribution 
of the matrix elements of the Wishart ensemble, in the case where
the numbers of rows and columns are finite and arbitrary. 
We gave a unified derivation which encompasses both the classical
Wishart case (where the number of rows is greater or equal to the number of
columns), and the opposite case, where part of the matrix is
necessarily nonrandom (anti-Wishart case). 

The expression obtained is a starting point for algorithms leading to 
fast reconstruction of the redundant information from the first 
$n$ rows of the Wishart matrix. The formulas for the reduced matrices
allow also to test an empirical Wishart-like matrix for redundant
information.
The more general problem of reconstructing  the
Wishart matrix from less regularly distributed  data points (sparse matrices) 
remains a challenging problem, which will be analyzed elsewhere.

\subsubsection*{Acknowledgments}
M.A.N. thanks Sergey Maslov and Yi-Cheng Zhang for discussions  
on  the Wishart-type matrices in relation 
to the redundant information problem, and for informing  about 
the  ongoing work on on real Anti-Wishart distribution~\cite{ZHANG2}.  
RJ thanks Yan Fyodorov for discussing his methods of~\cite{FYODOROV}. 
This work was supported in part by KBN grants 2P03B01917 and
2P03B09622. 

\medskip

\noindent{\bf Note added:} The number of constraints for the real 
Anti-Wishart
distribution is {\em the same} as in \cite{ZHANG2}. The apparent
discrepancy with the first version of this manuscript was only due to
a sloppy rewriting of a matrix valued Dirac delta function in terms of
component scalar Dirac delta functions.

\appendix

\section*{Appendix A -- Real Wishart and Anti-Wishart matrices}

In this appendix we will briefly summarize the relevant formulas for
$\OM=\AAd\AA$ where now $\AA$ is taken from a Gaussian {\em real} ensemble.
The determination of $P(\OM)$ follows the complex
case, using now the integral representation
\eq
\label{e.tint.r}
P_{n,k}(\OM)=\int d\TT \, \det(1+i\TT)^{-n/2} e^{i\tr \TT\OM}
\eqx
where we now integrate over {\em real symmetric} matrices $\TT$.
Instead of the integration by residues leading to (\ref{e.partial}),
we use the formula \cite{GR} 
\eq
\int_{-\infty}^{+\infty} dt \,
e^{itw}\frac{1}{(a+it)^{\nu}}=\frac{1}{\Gamma(\nu)} 2\pi w^{\nu-1}e^{-wa}
\eqx
Integrating over $t_{11}$ gives, modulo overall constants,
\eqn
\label{e.partial.r}
\int d\TT_{k-1} e^{i\tr \TT_{k-1}\OM_{k-1}} (\det(1+i\TT_{k-1}))^{-n/2} 
\om_{11}^{n/2-1}e^{-\om_{11}} \cdot \nonumber \\
\int dtdt^\dagger e^{-\om_{11} t^\dagger (1+i\TT_{k-1})^{-1} t}
e^{i (t^\dagger \om+\om^\dagger t)}
\eqnx
The last integral is real Gaussian giving 
\eq
\f{1}{\om_{11}^{\frac{k-1}{2}}}  \cdot [\det(1+i\TT_{k-1})]^{\frac{1}{2}} \cdot
e^{-\f{1}{\om_{11}} \om^\dagger (1+i\TT_{k-1}) \om}
\eqx 
Hence we are led to the recurrence relation
\eq
\label{e.rec.r}
P_{n,k}(\OM_k)\propto \om_{11}^{\frac{n-k-1}{2}} e^{-\om_{11}-\f{\om^\dagger
\om}{\om_{11}}} P_{n-1,k-1}\left(\OM_{k-1}- \f{1}{\om_{11}}\om
\om^\dagger \right)
\eqx

In the classical Wishart case (real variables) using
relation (\ref{e.rec.r}) repeatedly we may reduce the problem to
calculating $P_{n,1}(\OM)=\om^{n/2-1} e^{-\om}$. Again the recurrence
relation may be solved after diagonalizing $\OM$, and we recover the
result
\eq
P_{n,k}^{WISHART}(\OM)=C''''_{n,k}\, (\det\, \OM)^{\frac{n-k-1}{2}}\,
e^{-\tr\, \OM} 
\eqx
Similar iteration like in the  complex
Anti-Wishart case (19)  leads 
to matrix-valued delta for the reduced matrix $\Omega^{(k-n)}$.
It is defined by an analogous formula to (\ref{e.deltadef}) but with
the domain of integration being restricted to the space of {\em real
symmetric} $(k-n) \times (k-n)$ matrices. 
Its $(k-n)(k-n+1)/2$ independent elements can be
again rewritten  as  ratios of determinants, thus arriving at the
probability distribution 
\eq
P_{n,k}^{ANTI-WISHART}(\OM)=C''''_{n,k}\,
(\det\, \OM_{[n]})^{\frac{n-k-1}{2}}\,
e^{-\tr\, \OM}\, 
\prod_{l=n+1}^k \prod_{\stackrel{m=n+1}{l \leq m}}^k \dl\left( \f{\det\,
\OM_{[n],lm}}{\det\, \OM_{[n]}} \right)  
\eqx
The number of independent scalar deltas is equal to
$(k-n)(k-n+1)/2$ and agrees with the one in \cite{ZHANG2}.

For completeness, we remind  that 
joint eigenvalue distributions follow once we include the
appropriate real Vandermonde determinants:
\eqn
P^{WISHART}_{n,k}(\lm_1,\ldots,\lm_k)&=&\prod_i \lm_i^{\f{n-k-1}{2}} e^{-\lm_i}
\cdot \prod_{i<j} (\lm_i-\lm_j)^{\beta} \\
P^{ANTI-WISHART}_{n,k}(\lm_1,\ldots,\lm_n)&=&\prod_i \lm_i^{\f{k-n-1}{2}} e^{-\lm_i}
\cdot \prod_{i<j} (\lm_i-\lm_j)^{\beta} 
\eqnx
where $\beta=1$ for the real case. For recent results on  spectral analysis 
of the above formula we refer to~\cite{HAAG}. 

\section*{Appendix B -- Some details of the proof of (\ref{e.recsol}).}

The proof can be done by induction on $i$. For $i=0$ the statement is
obviously true. Since in this case we have
\eq
\OM^{(k)}_{lm}=\f{\det \OM_{[0],lm}}{\det \OM_{[0]}} =\OM_{lm}
\eqx
Let us assume that the formula (\ref{e.recsol}) holds for $i-1$. We
will verify the formula for $i$. To this end let us rewrite the
recurrence relation (\ref{e.matreceq})
\eq
\OM^{(k-i)}_{lm} = \OM^{(k-i+1)}_{k-i;lm} -
\f{\om^{(k-i+1)}\om^{(k-i+1)\dagger}}{\om^{(k-i+1)}_{11}} 
\eqx
as
\eq
\OM^{(k-i)}_{lm} = \OM^{(k-i+1)}_{l+1\, m+1}- \f{\OM^{(k-i+1)}_{l+1\,
1} \OM^{(k-i+1)}_{1\, m+1}}{\OM^{(k-i+1)}_{1\, 1}}
\eqx
where we made use of the decomposition (\ref{e.decomp}). We will now
insert the formula (\ref{e.recsol}) to obtain
\eq
\f{\det \OM_{[i]\, l+i\,m+i}}{\det \OM_{[i]}} =
\f{\det \OM_{[i-1]\, l+i\,m+i}}{\det \OM_{[i-1]}} -
\f{\det \OM_{[i-1]\, l+i\,i} \det \OM_{[i-1]\, i\,m+i}}{\det
\OM_{[i-1]} \det \OM_{[i-1]\, i\,i}}
\eqx
When we multiply this by $\det \OM_{[i]} \det \OM_{[i-1]}$ we see that
the induction step is equivalent to proving the following
determinental identity\footnote{Note that $\det \OM_{[i-1]i,i}=\det
\OM_{[i]}$.}: 
\eq
\label{e.identity}
\det \OM_{[i-1]} \det \OM_{[i]l+i\, m+i} = \det \OM_{[i-1]l+i\,m+i} \det
\OM_{[i-1]i\,i} - \det \OM_{[i-1]l+i\,i} \det \OM_{[i-1]i\,m+i}
\eqx
where we used the notation defined in (\ref{e.defi}) and
(\ref{e.defii}).
Now we use repeatedly, for each of the terms of (\ref{e.identity}) the
decomposition 
\eq
\det\left(
\begin{tabular}{ccc|c}
&          & &  \\
&$\OM_{[i-1]}$ & & $\psi$ \\
&          & &  \\
\hline
 &$\psi^\dagger$ & & $\sg$
\end{tabular}\right) =
\det \,\OM_{[i-1]} \cdot \left( \sg - \psi^\dagger \OM_{[i-1]}^{-1} \psi
\right)
\eqx
The only complication lies in applying this formula to $\det\,
\OM_{[i+1]}$ where terms involving $\OM_{[i]}^{-1}$ do appear. To
proceed we have to use a similar decomposition for the inverse matrix:
\eq
\OM_{[i]}^{-1} \equiv 
\left(
\begin{tabular}{ccc|c}
&          & &  \\
&$\OM_{[i-1]}$ & & $\psi$ \\
&          & &  \\
\hline
 &$\psi^\dagger$ & & $\sg$
\end{tabular}\right)^{-1} 
=
\left(
\begin{tabular}{ccc|c}
&          & &  \\
&$A$ & & $b$ \\
&          & &  \\
\hline
 &$b^\dagger$ & & $c$
\end{tabular}\right) 
\eqx
with
\eqn
A &=& \OM_{[i-1]}^{-1} - \, \OM_{[i-1]}^{-1} \psi b^\dagger \\
b &=& -c\, \OM_{[i-1]}^{-1} \psi \\
b^\dagger &=& -c\, \psi^\dagger \OM_{[i-1]}^{-1} \\
c &=& \f{1}{\sg -\psi^\dagger \OM_{[i-1]}^{-1} \psi}
\eqnx
After a straightforward but tedious calculation we arrive at
(\ref{e.identity}).

\end{document}